\documentclass[journal,twocolumn,10pt,final]{IEEEtran}

\usepackage[T1]{fontenc}
\usepackage{amsmath}
\usepackage{fancybox}

\usepackage{txfonts}

\newtheorem{definition}{Definition}
\newtheorem{proposition}[definition]{Proposition}
\newtheorem{theorem}[definition]{Theorem}

\newtheorem{remark}[definition]{Remark}
\newtheorem{lemma}[definition]{Lemma}

\begin{document}

\title{Universal Secure Multiplex Network Coding with Dependent and Non-Uniform
Messages}
\author{Ryutaroh~Matsumoto,~\IEEEmembership{Member, IEEE,} and 
Masahito Hayashi,~\IEEEmembership{Fellow, IEEE}%
\thanks{%
This research was partially supported by 
the MEXT Grant-in-Aid for Young Scientists (A) No.\ 20686026,
(B) No.\ 22760267, Grant-in-Aid for Scientific Research (A) No.\ 23246071,
and the Villum Foundation through their VELUX Visiting Professor Programme
2011--2012.
The Center for Quantum Technologies is funded
by the Singapore Ministry of Education and the National Research
Foundation as part of the Research Centres of Excellence programme.
This paper was
presented in part at 2011 IEEE International Symposium on Network Coding,
Beijing, China, July 2011
\cite{matsumotohayashi2011netcod}, and in part
at 2011 IEEE International Symposium on Information Theory, Saint
Petersburg, Russia, August 2011 \cite{matsumotohayashi2011isit}.}%
\thanks{%
R. Matsumoto was with Department of Information and Communications Engineering,
Tokyo Institute of Technology, 152-8550 Japan.
He is now with Department of Information and Communication Engineering,
Nagoya University, 464-8603 Japan
(email: ryutaroh@rmatsumoto.org).}%
\thanks{M. Hayashi is with
Graduate School of Mathematics,
Nagoya University, 464-8602 Japan, 
and
Centre for Quantum Technologies,
 National University of Singapore,
 3 Science Drive 2, Singapore 117542
 (email: masahito@math.nagoya-u.ac.jp).}
\thanks{Copyright \copyright\  2017 IEEE. Personal use of this material is permitted.  However, permission to use this material for any other purposes must be obtained from the IEEE by sending a request to pubs-permissions@ieee.org.
  Published version is available as
 \href{http://dx.doi.org/10.1109/TIT.2017.2694012}{DOI:10.1109/TIT.2017.2694012}}}
\date{April 13, 2017}

\allowdisplaybreaks
\maketitle

\begin{abstract}
We consider the random linear precoder at the source node
as a secure network coding.
We prove that it is strongly secure
in the sense of Harada and Yamamoto \cite{harada08}
and universal secure in the
sense of Silva and Kschischang \cite{silva09,silva08},
while allowing arbitrary small but nonzero mutual information to the
eavesdropper.
Our security proof allows statistically dependent
and non-uniform multiple secret messages,
while all previous constructions of weakly or strongly
secure network coding assumed independent and uniform
messages,
which are difficult to be ensured in practice.\\
\end{abstract}



\section{Introduction}\label{sec1}
Network coding \cite{ahlswede00} attracts much attention recently
because it can offer improvements in several metrics, such as throughput
and energy consumption, see \cite{fragouli07b,fragouli07a}.
On the other hand, the information theoretic security \cite{pls,liang09} also
attracts much attention  because it offers security that does not depend on
a conjectured difficulty of some computational problem.

A juncture of the network coding and the information theoretic security
is the secure network coding \cite{cai02b,caiyeung11}, which prevents an eavesdropper, called Eve,
from knowing the message from the legitimate sender,
called Alice, to the multiple legitimate receivers by eavesdropping intermediate links up to a specified number in a network.
In this paper, we focus on the single source multicast network coding. 
Here, we should remark that there are two kinds of formulation of (secure) network coding even in the single source multicast setting. 
In the first kind, given a graph corresponding to the network,
we design the coding operations
on each node to transmit information \cite{CF}.
In the second kind, given
(partial) information of the operations on intermediate nodes as well as the graph,
we design
the encoder and decoder on source and sink nodes, respectively.
We adopt the second formulation, and assume
linearity on the operations on intermediate nodes.

It can be seen \cite{rouayheb07,rouayheb12} as a network coding counterpart of the traditional wiretap channel coding problem considered by Wyner \cite{wyner75} and subsequently others \cite{liang09}.
In both secure network coding and coding for wiretap channels,
the secrecy is realized by including random bits
into the transmitted signal by Alice
so that the secret message
becomes ambiguous to Eve. The inclusion of random bits, of course,
decreases the information rate.
In order to get rid of the decrease in the information rate,
Yamamoto et~al.\ \cite{yamamoto05,kobayashi13}
proposed the secure multiplex coding for wiretap channels, in which
there is no loss of information rate.
The idea of Yamamoto et~al.\  is as follows:
Suppose that Alice has $T$ statistically independent messages
$S_1$, \ldots, $S_T$.
Then $S_1$, \ldots, $S_{i-1}$, $S_{i+1}$, \ldots,
$S_T$ serve as the random bits making $S_i$ ambiguous to Eve, for each $i$.
Indeed, since there are multiple legitimate receivers,
 each receiver may have a different demand for
 information. In this situation, it is natural that we have multiple messages $S_1$, \ldots, $S_T$ dependently on receivers' demands.

Independently and simultaneously,
Bhattad and Narayanan \cite{bhattad05} proposed
a scheme
based on the same idea as \cite{yamamoto05,kobayashi13},
whose goal is also to get rid of
the loss of information rate in the secure network coding.
This scheme was called weakly secure
  network coding in \cite{bhattad05}.
Their method \cite{bhattad05} ensures that the mutual information
between $S_i$ and Eve's information is zero for each $i$.
Recall that Eve's knowledge on secret information $S_i$ is
usually measured by the mutual information in
the information theoretic security \cite{pls,liang09}.
As drawbacks,
the construction depends on the network topology and coding at intermediate
nodes, and the computational complexity of code construction is large.

Harada and Yamamoto \cite{harada08} defined a stronger security
requirement on the weakly secure network coding,
which will be reviewed later, and called it as
the strongly
secure network coding.
Then they showed its construction procedure. As \cite{bhattad05},
the construction depends on the network topology and coding at intermediate
nodes, and the computational complexity of code construction is large.

In order to remove these drawbacks,
Silva and Kschischang \cite{silva09} proposed
a scheme called universal
weakly secure network coding, in which they showed an efficient
code construction that can support up to two $\mathbf{F}_q$-symbols in each $S_i$ and
is independent of the network topology and coding at intermediate
nodes, where $\mathbf{F}_q$ denotes the
finite field with $q$ elements throughout this paper.
The independence of coding at the source node from 
network topology and coding at intermediate
nodes is termed universal by Silva and Kschischang in \cite{silva09,silva08}.
They \cite{silva09} also  showed the existence of universal
weakly secure network coding with more than two $\mathbf{F}_q$-symbols
in $S_i$, but have not shown an explicit construction.

Cai \cite{cai09secure}
removed most of drawbacks mentioned earlier.
Cai proved that random linear network coding \cite{ho06}
gives the strongly
secure network coding in the sense of \cite{harada08} with
arbitrarily high probability with sufficiently large finite
fields. However,
he did not provide evaluation of the required field size, and
it seems huge. Moreover,
for some applications (e.g.\ \cite{yeung06b,yeung06a})
we want to choose coding at intermediate nodes
in non-random fashion.

There exists a common difficulty in all the previous constructions reviewed above. 
In practice, we are not sure if the multiple messages
are uniform and statistically independent. However,
all the previous studies\footnote{%
Cai \cite{cai09secure} considered arbitrary probability distribution
in \cite[Theorem 3.2]{cai09secure} but assumed
uniformity and independence for his study of the strongly
secure network coding in \cite[Section IV]{cai09secure}.}
assumed the uniformity and the independence,
and without both of them their security proofs do not seem to hold.
It is important to provide a security proof for weakly and strongly
secure network coding without uniformity or independence assumption.
On the other hand, non-uniformity of secret messages has been considered in
the ordinary secure network coding \cite{cai07securecondition,zhang09securecondition} (see also the survey \cite{cai11survey}).
In \cite{cai11survey,cai07securecondition,zhang09securecondition},
the randomness to hide a secret message was assumed to be statistically
independent of the secret message, while our present
study allows it to be
statistically dependent.

{We shall analyze the security of a slightly modified construction
of the random linear precoder originally proposed in \cite{cai02b}.}
Our modified construction is strongly secure in the sense of \cite{harada08}
and universal secure in the sense of \cite{silva09,silva08}.
{Uniformity and the independence assumptions are required in
previous works to guarantee security. This paper
relaxed the assumptions and aims to determine the
amount of information leakage if the two conditions are
not satisfied.}
The optimality of our modified construction is verified under the uniformity and
independence assumption at the end of
Remark \ref{rem:zeromutual}.

However, we relax an aspect of the security requirements
traditionally used in the secure network coding.
In previous proposals of secure network coding \cite{bhattad05,cai02b,harada08,silva09,silva08}
it is required that the mutual information to the eavesdropper is exactly
zero.
We relax this requirement by regarding sufficiently small mutual information
to be acceptable.
This relaxation is similar to requiring the
decoding error probability
to be sufficiently small instead of strictly zero.
Also observe that our relaxed criterion is much stronger than
one commonly used in the information theoretic security \cite{liang09}.
Our modified construction can realize arbitrary small mutual information
if coding over sufficiently many symbols in single packet is allowed.

Up to this point,
we have followed the conventional usage of terminology
``strong security'' and ``weak security'' in secure network coding.
On the other hand, in the context of key agreement and
wiretap channel coding and ``strong security'' and ``weak security''
mean completely different security criteria \cite{bloch13}.
We shall introduce a different
terminology ``secure multiplex network coding'' to mean
``strong security'' used in secure network coding.

After we submitted the original manuscript in 2012,
one of the authors started and published another approach
\cite{kurihara15} to the same problem as this paper.
\cite{kurihara15} proposed a deterministic construction
of universal secure multiplex network coding
and its security analysis also valid for dependent and
non-uniform multiple messages, while the proposed
construction in this paper is probabilistic.
However, when multiple messages are dependent or
non-uniform, the construction and the security analysis
in \cite{kurihara15} cannot ensure the mutual information to
the eavesdropper arbitrarily small, which makes
the construction in \cite{kurihara15} less useful for
dependent or non-uniform messages.
As far as the authors know,
only the construction in the present paper can
ensure arbitrarily small mutual information to the eavesdropper
when multiple messages are dependent or
non-uniform.

This paper is organized as follows:
Section \ref{sec2} reviews related results used in this paper,
and a slightly new terminology ``secure multiplex network coding''.
Section \ref{sec3} introduces the strengthened version of the privacy
amplification theorem and the proposed scheme
for secure network coding.
Section \ref{sec4} concludes the paper.

Part of this paper was reported as earlier proceedings papers
\cite{matsumotohayashi2011isit,matsumotohayashi2011netcod}.
We substantially rewrote our security proof in \cite{matsumotohayashi2011netcod}
so that we can analyze the security with dependent and non-uniform multiple
secret messages, which was not done in \cite{matsumotohayashi2011netcod}.
We borrowed ideas from \cite[Section IV]{matsumotohayashi2011isit}
and extended them in Appendix \ref{app:b}
so that we can prove Lemma \ref{lem10}.

\section{Preliminary}\label{sec2}
\subsection{Model of network and network coding
and two-universal hash functions}\label{sec21}
As in \cite{bhattad05,cai02b,caiyeung11,harada08,silva09,silva08}
we consider the single source multicast, and
assume the linear network coding \cite{koetter03,li03}.
The source node is assumed to have  at least $n$ outgoing links.
For $i=1$, \ldots, $n$, the source node generates a packet $P_i$
consisting of $m$ symbols in $\mathbf{F}_q$,
and transmits an $\mathbf{F}_q$-linear combination of $P_1$,
\ldots, $P_n$ to each outgoing link, as explained in \cite[Section 2.1]{fragouli06ccr}.
At an intermediate node, only packets generated at the same time
by the source node
are linearly combined, as explained in \cite[Section 2.5]{fragouli06ccr}.
The linear combination coefficients at each node are fixed so that
all the legitimate receivers can decode $n$ packets
$P_1$, \ldots, $P_n$ from the source node.
In this paper, we assume that all of sink nodes have respective decoders to recover all of the $nm$  transmitted symbols.  
Since all of legitimate receivers can recover the message without error due to this assumption, 
we do not need to discuss the decoding error probability, and focus on the security.

If the random linear network coding \cite{ho06} is employed,
we have to also include so-called encoding vectors in
each packet $P_i$ \cite[Section 2.2]{fragouli06ccr}.
We ignore those encoding vectors because they do not carry
secret information.

Hereafter, we shall only consider the eavesdropper Eve and
forget about the multiple legitimate receivers.
The $n$ packets $P_1$, \ldots, $P_n$ carry in total
$mn$ symbols in $\mathbf{F}_q$.
We shall propose a method encoding secret information into
$mn$ symbols by
the source node. The $mn$ symbols obtained by the proposed method
are distributed to packets $P_1$, \ldots, $P_n$.

Eve can eavesdrop $\mu$ links.
We assume $\mu \leq n$ throughout this paper.
The total number of eavesdropped symbols is therefore $m \mu$.
The set of $\mu$ eavesdropped links is assumed to be fixed during
packets $P_1$, \ldots, $P_n$ are traveling on the network,
as assumed in \cite{silva09,silva08}.
The situation considered here also includes the conventional
store-and-forward network as a special case.

We shall use a family of two-universal hash functions \cite{carter79}
for the privacy amplification theorem introduced later.
\begin{definition}\label{def:twouniv}
Let $\mathcal{F}$ be a set of functions from a finite set $\mathcal{S}_1$ to
another finite  set $\mathcal{S}_2$,
and $F$ a random variable on $\mathcal{F}$. If for any $x_1 \neq x_2
\in \mathcal{S}_1$ we have
\begin{equation}
\mathrm{Pr}[F(x_1)=F(x_2)] \leq \frac{1}{|\mathcal{S}_2|}, \label{eq:two}
\end{equation}
then $\mathcal{F}$ with the probability distribution of
$F$ is said to be a \emph{family of two-universal hash functions}.
\end{definition}


\subsection{Security definitions}
In this subsection,
we  review the existing security criteria,
and introduce our security criterion.
We also discuss the relation among security criteria
because the same terminology is used to mean different criteria.

\begin{definition}[Strongly secure network coding]
\cite{harada08}
Let $m=1$, and $S_1$, \ldots, $S_T \in \mathbf{F}_q$
be messages with $T \leq n$.
{We denote by $S_{T+1}$, \ldots, $S_n \in \mathbf{F}_q$
randomness not intended as messages.}
A network coding is said to be $\eta$-strongly secure if
the following relation holds for any $0 \leq \mu \leq n$.
When Eve's observation $Z$ is obtained by eavesdropping $\mu$ links,
any $\mathcal{I} \subset \{1$, \ldots, $T\}$
with $\mu-\eta \leq T-|\mathcal{I}|$
satisfies 
\[
I(S_{\mathcal{I}};Z ) = 0,
\]
where $S_{\mathcal{I}} = [S_i : i \in \mathcal{I}]$ and
$I(S_{\mathcal{I}};Z)$ denotes their  mutual information
as defined in \cite{cover06}.
\end{definition}

{The parameter $\eta$ is equivalent to $k$ in \cite{harada08}.}
Harada and Yamamoto \cite{harada08} showed a procedure
to construct $(n-T)$-strongly secure network coding
under the uniformity and independence assumption on the messages
$S_1$, \ldots, $S_n$.
Bhattad and Narayanan \cite{bhattad05} introduced the weak security for
network coding that requires $I(S_{i};Z ) = 0$ for all $i \in \mathcal{I}$.

We want to consider the universal security studied in \cite{silva09,silva08},
and also want to use multiple symbols in a single packet $P_i$,
that is,  $m > 1$.
So we introduce our version of universal strong security,
by following the approach initiated by Silva and Kschischang \cite{silva09,silva08}.

\begin{definition}\label{def:univstrongsec}
Assume that we are given a linear network coding
for single source multicast.
Assume also that linear coding at intermediate nodes and
the set of $\mu$ eavesdropped links are fixed when packets $P_1$, \ldots,
$P_n$ travel from the source node to all the legitimate receivers.
Suppose that we have $T+1$ messages $S_1$, \ldots, $S_{T+1}$ and
$S_i \in \mathbf{F}_q^{k_i}$. 
$S_{T+1}$ denotes randomness not intended as a message.
We assume $\sum_{i=1}^{T+1} k_i = mn$.
A linear transformation of $S_1$, \ldots, $S_{T+1}$ at the
source node is said to be a universal $(\epsilon,\eta)$-secure multiplex network coding 
if the following relation holds 
for all linear coding at intermediate nodes
and for any $0 \leq \mu \leq n$. 
When Eve's observation $Z$ corresponds to $\mu$ eavesdropped links, 
any subset $\mathcal{I} \subset \{1$, \ldots, $T\}$
with $m(\mu-\eta) < \sum_{1\leq i \leq T+1, i \notin \mathcal{I}} k_i$
satisfies
\begin{equation}
I(S_{\mathcal{I}};Z )
\leq \epsilon,\label{eq:strongsecure}
\end{equation}
where
$S_{\mathcal{I}} = [S_i : i \in \mathcal{I}]$.
\end{definition}

Readers may observed that the above secure
  multiplex network coding
with $\epsilon=0$ is almost the same as
the strong security in \cite{harada08}.
The reason for using a different name is as follows.
In the study of wiretap channel coding,
we usually consider a sequence of encoders and decoders
for block length $m=1$, $2$, \ldots.
the weak security in the wiretap coding means 
$\lim_{m\rightarrow\infty} I(S,Z)/m=0$,
where $S$ is the message of the wiretap coding and $Z$ is the
received sequence by the eavesdropper.
The strong security in the wiretap coding means
$\lim_{m\rightarrow\infty} I(S,Z)=0$.
Since those meanings of the weak and strong security in the wiretap
coding are different from the secure network coding,
we introduced a 
different terminology in Definition \ref{def:univstrongsec}
to reduce unnecessary confusion.

\section{Universal secure multiplex network coding}\label{sec3}
\subsection{Strengthened privacy amplification theorem}
In order to evaluate the mutual information to Eve when the
sum rate of multiple secret information is large,
we need to strengthen the privacy amplification theorem
originally appeared in \cite{bennett95privacy,hayashi11}
as follows.
The below new privacy amplification theorem
enables an upper bound (\ref{eq:ub7}) on the mutual information
when the mutual information grows with $m$ instead of
converging to zero.

The following proposition is a slightly enhanced version of 
\cite[Theorem 2]{matsumotohayashi2011netcod}.
\begin{proposition}\label{thm2}
Let $A_1$ and $A_2$ be discrete random variables on finite sets $\mathcal{A}_1$ and
$\mathcal{A}$, respectively, and
$\mathcal{F}$ a family of functions from $\mathcal{A}_1$ to $\mathcal{A}_3$.
Let $F$ be a random variable on $\mathcal{F}$.  
Assume that $A_1$ and $F$ are conditionally independent given $A_2$,
and that for any fixed realization $a_2$ of $A_2$, the
conditional probability distribution
of $F$ given $a_2$ satisfies the condition for a family of two-universal
hash functions.
Then we have
\begin{equation}
\mathbf{E}_f [\exp(\rho I(F(A_1);A_2|F=f))]  \leq   
1+ |\mathcal{A}_3|^\rho\mathbf{E}[P_{A_1|A_2}(A_1|A_2)^\rho]\label{hpa1}
\end{equation}
for all $0 \leq \rho \leq 1$,
{where $\mathbf{E}_f[ \cdot ]$ denotes the expectation of $\cdot$ with
$f$ being the random variable.}
We use the natural logarithm for
all the logarithms in this paper,
which include ones implicitly appearing in entropy and mutual information.
Otherwise we have to adjust the above inequality.
\end{proposition}
\begin{IEEEproof}
Proof is given in Appendix \ref{app:a}.
\end{IEEEproof}
{In our analysis of the security,
we shall use Proposition \ref{thm2} with
$A_1$ being the whole secret message,
$A_2$ being part of the secret message whose secrecy we analyze,
and $F(A_1)$ being Eve's observation.}

\subsection{Description of the proposed scheme and 
analysis with randomized coding}\label{sec31}
The purpose of this section is
to provide a universal
$(\epsilon_I, (k_{T+1}/m - \delta_\rho))$-secure multiplex
network coding in the sense of Definition \ref{def:univstrongsec}, where $\delta_\rho$ is 
a parameter measuring conditional non-uniformity
to be defined in Eq.\ (\ref{eq:delta0}).
The modified sense means that the zero mutual information
in Eq.\ (\ref{eq:strongsecure}) is
relaxed to the requirement that it can be made arbitrarily small.
For this purpose, in this subsection,
we treat the coding scheme with randomized coding.
We assume that we have $T$ secret messages, which can be
dependent or non-uniform, and
that the $i$-th secret message is given as a random variable $S_i$
whose realization is a row vector in $\mathbf{F}_q^{k_i}$.
We shall provide upper bounds on the information leaked to
Eve for all choices of values of $k_i$.
We shall also use a supplementary random message
$S_{T+1}$ taking values in $\mathbf{F}_q^{k_{T+1}}$
when the randomness in the encoder is insufficient to make
$S_i$ secret from Eve.
By $S$ we denote the entire collection $(S_1$, \ldots, $S_{T+1})$
of messages.
We assume $mn= k_1 + \cdots+k_{T+1}$.

\begin{figure*}[t!]
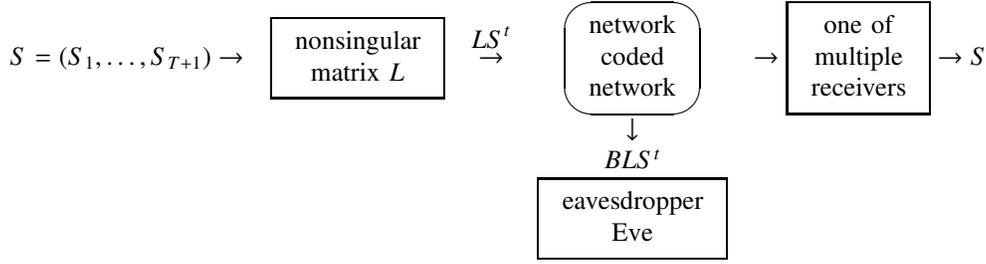

\[
\begin{array}{ccccc}
S=(S_1,\ldots, S_{T+1}) \rightarrow &\fbox{\begin{tabular}{c}nonsingular\\ matrix $L$\end{tabular}}& \overset{\displaystyle LS^t}{\rightarrow} &\ovalbox{\begin{tabular}{c}network\\
coded\\ network\end{tabular}}& \rightarrow\fbox{\begin{tabular}{c}one of \\ multiple\\ receivers\end{tabular}}\rightarrow S\\
&&&\downarrow&\\
&&&BLS^t&\\
&&&\fbox{\begin{tabular}{c}eavesdropper\\ Eve\end{tabular}}&
\end{array}
\]
\caption{Proposed coding scheme for the universal secure multiplex network coding}\label{fig1}
\end{figure*}

Let $\mathcal{L}$ be the set of all bijective $\mathbf{F}_q$-linear
maps from $\prod_{i=1}^{T+1} \mathbf{F}_q^{k_i}$ to itself,
and $L$ the uniform random variable on $\mathcal{L}$
statistically independent of $S=(S_1$, \ldots, $S_{T+1})$,
and arbitrary fix nonempty $\mathcal{I} \subseteq \{1$, \ldots, $T\}$.
The source node store $LS^t$ into packets $P_1$, \ldots, $P_n$
defined in Section \ref{sec21} and send them via its $n$ outgoing links,
where $t$ denotes the transpose of a vector.
Our modified construction just
adds a bijective linear precode to an existing network code.
{Note that attaching a random linear function was first proposed in
\cite{cai02b} for the secure network coding.}
This coding scheme is illustrated in Fig.\ \ref{fig1}.

The legitimate sender and all the legitimate receivers agree on
the choice of $L$. The eavesdropper
Eve may also know their choice of $L$.
Choice of $L$ is part of protocol specification,
the chosen $L$ is repeatedly used,
and agreement on its choice among legitimate sender and receivers
is not counted as consumption of the network bandwidth.
A legitimate receiver can recover $S_1$, \ldots, $S_T$,
$S_{T+1}$ by multiplying $L^{-1}$ to his/her received information.
By the assumption on Eve, her information can be expressed as
$BLS^t$ by using an $m\mu  \times mn$ matrix $B$ over $\mathbf{F}_q$
as in \cite{silva09,silva08}.

For the nonempty $\mathcal{I} \subseteq \{1$, \ldots, $T\}$,
denote the collection of random variables $[S_i : i \in \mathcal{I}]$
by $S_{\mathcal{I}}$, denote
$[S_i : i \in \{1$, \ldots, $T+1\}\setminus \mathcal{I}]$
by $S_{\overline{\mathcal{I}}}$,
and let
$k_{\mathcal{I}} = \sum_{i\in\mathcal{I}} k_i$.

For a fixed realization $\ell$ of $L$,
the information gained by Eve is measured by
the mutual information $I(S_{\mathcal{I}};BLS^t|L=\ell)$,
which is common practice in the information theoretic security
\cite{pls,liang09}.
Since its average $\mathbf{E}_\ell[
I(S_{\mathcal{I}};BLS^t|L=\ell)]$ is the conditional mutual
information $I(S_{\mathcal{I}};BLS^t|L)$ \cite{cover06},
we will  upper bound $I(S_{\mathcal{I}};BLS^t|L)$.
After upper bounding the average $I(S_{\mathcal{I}};BLS^t|L)$
in Eq.\ (\ref{eq1003}),
we can ensure that for most choices of $\ell$ and all possible
$B$,
$I(S_{\mathcal{I}};BLS^t|L=\ell)$ is small, as done in Eq.\ (\ref{eq:prob}).

In order to use Proposition \ref{thm2},
we introduce a lemma.
\begin{lemma}\label{lem10}
For fixed $B$,
the family of mapping $S \mapsto
BLS^t$ is a family of two-universal hash functions
to the $\mathrm{rank}(B)$-dimensional $\mathbf{F}_q$-linear space.
\end{lemma}
\begin{IEEEproof}
See Appendix \ref{app:b}.
\end{IEEEproof}

We can  upper bound
$I(S_{\mathcal{I}};BLS^t|L)$ as follows,
by applying Proposition \ref{thm2} with $A_1=S$,
$A_2=S_{\mathcal{I}}$, and $F(A_1)=BLS^t$.
{Observe that the assumption in Proposition \ref{thm2}
holds because $S_{\mathcal{I}}$ is part of $S$ and $L$ is independent of $S$.}
\begin{eqnarray}
&& \mathbf{E}_\ell [\exp(\rho I(S_{\mathcal{I}};BLS^t|L=\ell))]\nonumber\\
&\leq& 1+ q^{m \rho \times \mathrm{rank}(B)}\mathbf{E}[P_{S|S_{\mathcal{I}}}(S|S_{\mathcal{I}})^\rho]\nonumber\\
&=& 1+ q^{m \rho \times \mathrm{rank}(B)}\mathbf{E}[P_{S_{\overline{\mathcal{I}}}|S_{\mathcal{I}}}(S_{\overline{\mathcal{I}}}|S_{\mathcal{I}})^\rho]\nonumber\\
&\leq& 1+ q^{m \rho \mu}\mathbf{E}[P_{S_{\overline{\mathcal{I}}}|S_{\mathcal{I}}}(S_{\overline{\mathcal{I}}}|S_{\mathcal{I}})^\rho].\label{eq1002}
\end{eqnarray}From Eq.\ (\ref{eq1002})
we have
\begin{eqnarray}
&& \rho I(S_{\mathcal{I}};BLS^t|L)\nonumber\\
&=&\ln\exp (\rho I(S_{\mathcal{I}};BLS^t|L))\nonumber\\
&\leq&\ln \mathbf{E}_\ell [\exp(\rho I(S_{\mathcal{I}};BLS^t|L=\ell))]\nonumber\\
&\leq&\ln (1+ q^{m \rho \mu}\mathbf{E}[P_{S_{\overline{\mathcal{I}}}|S_{\mathcal{I}}}(S_{\overline{\mathcal{I}}}|S_{\mathcal{I}})^\rho])\nonumber\\
&\leq&q^{m \rho \mu}\mathbf{E}[P_{S_{\overline{\mathcal{I}}}|S_{\mathcal{I}}}(S_{\overline{\mathcal{I}}}|S_{\mathcal{I}})^\rho].\label{eq1003}
\end{eqnarray}

Fix  a real number $C_1 > 1$.
Equation (\ref{eq1003}) and the Markov inequality yield that
\begin{align*}
\mathrm{Pr} [\ell\in \mathcal{L}_{\mathcal{I},1}] <  1/C_1 
\end{align*}
for any single nonempty $\mathcal{I} \subseteq \{1$, \ldots, $T\}$,
where $\mathcal{L}_{\mathcal{I},1}$ $:=$
$\{ \ell \mid  I(S_{\mathcal{I}};BLS^t|L=\ell) > C_1 \mathbf{E}_{\ell} [I(S_{\mathcal{I}};BLS^t|L=\ell)] \}$.
Thus,
\[
\mathrm{Pr} [\ell \in \cup_{\mathcal{I}: \mathcal{I}\neq \emptyset} \mathcal{L}_{\mathcal{I},1}]
 <   (2^T-1)/C_1.
\]
This means that there is at least a probability of
 $1-(2^T-1)/C_1$ such that a realization $\ell$ of $L$
satisfies 
\begin{align}
&  I(S_{\mathcal{I}};BLS^t|L=\ell) \nonumber \\
& \leq 
C_1 \mathbf{E}_{\ell} [I(S_{\mathcal{I}};BLS^t|L=\ell)] \nonumber \\
& \leq 
C_1 q^{m \rho \mu}\mathbf{E}[P_{S_{\overline{\mathcal{I}}}|S_{\mathcal{I}}}(S_{\overline{\mathcal{I}}}|S_{\mathcal{I}})^\rho]/\rho\label{eq:ub5}
\end{align}
for all the $(2^T-1)$ nonempty subsets $\mathcal{I}$ of $\{1$, \ldots, $T\}$.
Defining another subset $\mathcal{L}_{\mathcal{I},2}$ $:=$
$\{ \ell \mid 
\exp(\rho I(S_{\mathcal{I}};BLS^t|L=\ell))
> 
C_1 
\mathbf{E}_{\ell} [\exp(\rho I(S_{\mathcal{I}};BLS^t|L=\ell))]\}$,
by Eq.\ (\ref{eq1002}) and the Markov inequality we obtain 
\[
\mathrm{Pr} [\ell \in \cup_{\mathcal{I}: \mathcal{I}\neq \emptyset} 
(\mathcal{L}_{\mathcal{I},1} \cup \mathcal{L}_{\mathcal{I},2})]
 <   2 (2^T-1)/C_1.
\]
Therefore, 
a realization $\ell$ of $L$ satisfies both Eq. (\ref{eq:ub5}) and
\begin{align}
 \exp(\rho I(S_{\mathcal{I}};BLS^t|L=\ell))
&\leq  C_1 (1+ q^{m \rho \mu}\mathbf{E}[P_{S_{\overline{\mathcal{I}}}|S_{\mathcal{I}}}(S_{\overline{\mathcal{I}}}|S_{\mathcal{I}})^\rho]).\label{eq:ub6}
\end{align}
with probability  at least $1-2 \times (2^T-1)/C_1$.

Equation (\ref{eq:ub6}) implies 
\begin{align}
& \frac{I(S_{\mathcal{I}};BLS^t|L=\ell)}{m}\nonumber\\
& = \frac{1}{m}\ln \exp I(S_{\mathcal{I}};BLS^t|L=\ell)\nonumber\\
&\leq  \frac{\ln C_1}{m\rho} + \frac{1}{m\rho}\ln
(1+q^{m \rho \mu}\mathbf{E}[P_{S_{\overline{\mathcal{I}}}|S_{\mathcal{I}}}(S_{\overline{\mathcal{I}}}|S_{\mathcal{I}})^\rho]) \textrm{ (by Eq.\ (\ref{eq:ub6}))}\nonumber\\
&\leq
 \frac{\ln C_1}{m\rho} + 
\left|\mu \ln q + \frac{1+\ln \mathbf{E}[P_{S_{\overline{\mathcal{I}}}|S_{\mathcal{I}}}(S_{\overline{\mathcal{I}}}|S_{\mathcal{I}})^\rho]}{m\rho}\right|^+, \label{eq:ub7}
\end{align}
where in Eq.\ (\ref{eq:ub7}) we used $\ln(1+\exp(x)) \leq |1+x|^+ = \max\{0$,
$1+x\}$.

Summarizing the preceding discussion, we have the following proposition.
\begin{proposition}\label{lem:premain}
Recall  that the eavesdropping $m\mu  \times mn$ matrix $B$ is fixed,
that $L$ is the uniform random variable on $\mathcal{L}$
statistically independent of $S=(S_1$, \ldots, $S_{T+1})$,
and that a real number $C_1 > 1$ is arbitrarily fixed.
There is at least a probability of $1-2 \times (2^T-1)/C_1$
such that information leakage $I(S_{\mathcal{I}};BLS^t|L=\ell)$
to Eve with the chosen realization $\ell$ of $L$
satisfies both inequalities (\ref{eq:ub5}) and (\ref{eq:ub7})
simultaneously.
\end{proposition}

The previous proposition does not ensure
the universal security in the sense of
\cite{silva09,silva08}
because it only considers a fixed eavesdropping matrix $B$.
To ensure the universal security, we must consider
all the possible eavesdropping matrix $B$, which shall be
done in the next two subsections.

\subsection{Evaluation of the number of different kinds of eavesdropping}\label{sec33}
In the following,
we considered the case when 
the matrix $B$ corresponds to $\mu$ eavesdropped links.
Such a case can be mathematically formulated as follows.
Let $x_{i,j} \in \mathbf{F}_q$ be the $j$-th symbol in the $i$-th
packet $P_i$ defined in Section \ref{sec21}.
Then there exists a $\mu \times n$ matrix $B_{\mu \times n}$ such
that what are observed by Eve at the $j$-th symbols
in her eavesdropped $\mu$ packets is expressed as
$B_{\mu \times n} ( x_{1,j}$, \ldots, $x_{n,j})^t$ for $j=1$, \ldots, $m$.
Without loss of generality we may assume $\mathrm{rank}(B_{\mu \times n}) = \mu$ because if $\mathrm{rank}(B_{\mu \times n}) = \mu' < \mu$ then such a
case can be regarded as only $\mu'$ links being eavesdropped.
Then,
the $m\mu  \times mn$ matrix\footnote{Mathematically,
the $m\mu  \times mn$ matrix $B$ is written as
$B_{\mu \times n} \otimes I_{m \times m}$.} $B$ is completely determined by $B_{\mu \times n}$.

In order to show the universal security in 
Definition \ref{def:univstrongsec},
we need to ensure that the mutual information is small for any $B$
and any $0 \leq \mu \leq n$.
For this purpose,
we need to count the number of different kinds of eavesdropping.

We consider the set $\mathcal{B}(\mu)$ of all possible $m\mu  \times mn$
matrices $B$ that
characterize Eve's eavesdropping with the above restriction.
Then, we define an equivalence relation $\sim$
on $\mathcal{B}(\mu)$ as $B_1 \sim B_2$ for $B_1, B_2 \in \mathcal{B}(\mu)$
if there exists an invertible function $f$ such that
$f (B_1 LS^t)= B_2 LS^t$ for all $L$ and $S^t$.
That is, $B_1 \sim B_2$ if and only if the
kernel of $B_1$ is the same as that of $B_2$.
Since $B_1$ and $B_2$ are determined by $\mu  \times n$ matrices,
the space $\mathcal{B}(\mu)/\sim $ is the set of the
$(n-\mu)$-dimensional subspaces in 
$\mathbf{F}_q^n$.
The space is called Grassmannian 
and the number is evaluated in the following way \cite{Ex}
\begin{align}
& |\mathcal{B}(\mu)/\sim| 
=
\prod_{i=0}^{\mu-1}
\frac{q^{n}-q^i}{q^{\mu}-q^i}
\leq 
\prod_{i=0}^{\mu-1}
\frac{q^{n}-q^{\mu-1}}{q^{\mu}-q^{\mu-1}}
= 
\prod_{i=0}^{\mu-1}
\frac{q^{n-\mu+1}-1}{q-1} \nonumber \\
\le &
\prod_{i=0}^{\mu-1}
q^{n-\mu+1}
= q^{\mu (n-\mu+1)}
\le
q^{\frac{(n+1)^2}{4}}
\label{h-1}
\end{align}
because $(x-z)/(y-z)$ is monotonically increasing in $z$ when $x>y>z>0$.
The final inequality follows from the inequality
$\sqrt{\mu (n-\mu+1)} \le \frac{\mu+n-\mu+1}{2}=\frac{n+1}{2}$.
Hence,
the total number of equivalence classes
excluding $B(0)$ is upper bounded as
\begin{align}
\sum_{\mu=1}^n |\mathcal{B}(\mu)/\sim| & \leq 
n q^{\frac{(n+1)^2}{4}}.
\label{eq-9-14}
\end{align}

\subsection{Universally secure multiplex network coding}
Next, using the above discussion,
we show the existence of universal
secure multiplex networking coding.
Due to (\ref{eq-9-14}),
the probability of $L$ satisfying Eqs. (\ref{eq:ub5}) and
(\ref{eq:ub7}) simultaneously for all possible $B$ is at least
\begin{equation}
1-2 \times (2^T-1)\times  n q^{\frac{(n+1)^2}{4}} /C_1. \label{eq:prob}
\end{equation}
Recall that chosen $L$ is part of protocol specification and
repeatedly used.
Because Eqs. (\ref{eq:ub5}), (\ref{eq:ub7}) and (\ref{eq:prob})
are independent of realization of the random variable $S$ representing
secret information, Eqs. (\ref{eq:ub5}) and (\ref{eq:ub7}) are satisfied
in every repeated use of $L$ with probability
at least Eq.\ (\ref{eq:prob}).

The upper bound (\ref{eq:ub5}) can go to either zero or $\infty$
as $m\rightarrow \infty$.
When the upper bound (\ref{eq:ub5}) goes to $\infty$,
the information leakage to Eve grows linearly with $m$ and
its growth rate with $m$ will be analyzed by Eq. (\ref{eq:ub7}).
Firstly, we need to clarify under what condition Eq. (\ref{eq:ub5})
converges to zero as $m \rightarrow \infty$.
To do so, we shall introduce a version of conditional R\'enyi entropy
introduced in \cite{hayashi11}.
There seems to be no standard definition for the 
conditional R\'enyi entropy, for example, definitions in
\cite{bennett95privacy} and \cite{golshani09} disagree and
our definition in \cite{hayashi11} is different from
\cite{bennett95privacy,golshani09}.
For discrete random variables $X$, $Y$,
define conditional R\'enyi entropy of order $1+\rho$ as
\[
H_{1+\rho}(X|Y) = -\frac{\ln \mathbf{E}[P_{X|Y}(X|Y)^\rho]}{\rho}.
\]
For $\rho=0$, we define $H_1(X|Y)$ as $\lim_{\rho\rightarrow 0}
H_{1+\rho}(X|Y)$. 
By using l'H\^opital's rule we see that $H_1(X|Y)$ is equal to the
conditional Shannon entropy.
Observe also that $H_{1+\rho}(X|Y) = \log_q |\mathcal{X}|$
if $X$ is conditionally uniform given $Y$, where
$\mathcal{X}$ denotes the alphabet of $X$.
We note that
$\mathbf{E}[P_{S_{\overline{\mathcal{I}}}|S_{\mathcal{I}}}(S_{\overline{\mathcal{I}}}|S_{\mathcal{I}})^\rho] = e^{H_{1+\rho}(S_{\overline{\mathcal{I}}}|S_{\mathcal{I}})}$.

In order to
clarify under what condition Eq. (\ref{eq:ub5})
converges to zero,
we need to assume some knowledge on $P_{S_{\overline{\mathcal{I}}}|S_{\mathcal{I}}}(S_{\overline{\mathcal{I}}}|S_{\mathcal{I}})$.
We consider the situation in which each message $S_i$ originates from
a different organization and it is compressed before network coded.
Even after compression, it is known that
$S_1$, \dots, $S_T$ are not completely uniform \cite{hayashi08},
and we must allow certain degree of statistical dependence among
$S_1$, \ldots, $S_T$ and their non-uniformity.
In this paper we consider secure network coding separately from
source coding of $S_i$.

Let $\delta_\rho$ be a nonnegative constant
such that
\begin{equation}
n-\frac{k_{\mathcal{I}}}{m} - \frac{H_{1+\rho}(S_{\overline{\mathcal{I}}}|S_{\mathcal{I}})}{m\ln q} \leq \delta_\rho \label{eq:delta0}
\end{equation}
for some $0< \rho \leq 1$, for all $\mathcal{I}$, and for sufficiently large $m$.
Observe that if all messages $S_i$'s are uniform and independent
then $\delta_\rho=0$. The parameter $\delta_\rho$ captures the deviation from the
uniform and independent situation in terms of conditional R\'enyi entropy
per the number $m$ of symbols in single packet.
By taking the natural logarithm of Eq.\ (\ref{eq:ub5}), we see
\begin{eqnarray}
&&  \ln \mbox{[RHS of Eq.\ (\ref{eq:ub5})]}\nonumber\\
&=& \ln \frac{C_1}{\rho} +m\rho (\mu\ln q +\frac{\ln\mathbf{E}[P_{S_{\overline{\mathcal{I}}}|S_{\mathcal{I}}}(S_{\overline{\mathcal{I}}}|S_{\mathcal{I}})^\rho]}{m\rho}) \nonumber\\
&=& \ln \frac{C_1}{\rho} +m\rho (\overbrace{\mu-\frac{H_{1+\rho}(S_{\overline{\mathcal{I}}}|S_{\mathcal{I}})}{m}}^{(*)})\ln q .
 \label{eq1004}
\end{eqnarray}
When 
\begin{equation}
\mu < (n-\frac{k_{\mathcal{I}}}{m}) - \delta_\rho
\mbox{ i.e.\ } \frac{k_{\mathcal{I}}}{m} < n-\mu-\delta_\rho,\label{zerocond}
\end{equation}
$(*)$ in Eq.\ (\ref{eq1004})
becomes negative by Eq.\ (\ref{eq:delta0}).
Under such condition Eq.\ (\ref{eq1004}) converges to $-\infty$
as $m\rightarrow \infty$, which means
that the upper bound Eq.\ (\ref{eq:ub5}) can be made arbitrary small
by letting $m$ be large.

Secondly, we shall analyze how much information Eve can gain when
Eq.\ (\ref{zerocond}) does not hold.
In such case we use the other upper bound Eq.\ (\ref{eq:ub7}).
We can rewrite Eq.\ (\ref{eq:ub7}) as
\begin{eqnarray*}
&& \mbox{RHS of Eq.\ (\ref{eq:ub7})}\\
&=& \frac{1+\ln C_1}{m\rho} + 
\mu \ln q - \frac{H_{1+\rho}(S_{\overline{\mathcal{I}}}|S_{\mathcal{I}})}{m}\\
&\leq& 
\frac{1+\ln C_1}{m\rho} + (\mu - (n-\frac{k_{\mathcal{I}}}{m}-\delta_\rho))\ln q
\mbox{ (by Eq.\ (\ref{eq:delta0}))}.
\end{eqnarray*}
We see that  we can make the upper bound
Eq.\ (\ref{eq:ub7}) on
$\frac{I(S_{\mathcal{I}};BLS^t|L=\ell)}{m}$ arbitrary close
to 
\begin{equation}
(\mu + \delta_\rho - (n - \frac{k_{\mathcal{I}}}{m}))\ln q \label{eq1005}
\end{equation}
by letting $m$ be large.

Observe that the assumption (\ref{zerocond})
is equivalent to the assumption of Definition \ref{def:univstrongsec}
with $\eta = k_{T+1}/m - \delta_\rho$.
By summarizing the previous discussion,
we can construct a universal secure multiplex
network coding in the sense of
Definition \ref{def:univstrongsec} as follows:
\begin{theorem}\label{th:main}
For any $\epsilon_p, \epsilon_I > 0$ and sufficiently large $m$,
a random choice of $mn\times mn$ matrix $L$ gives with probability at least $1-\epsilon_p$
a universal $(\epsilon_I$, $k_{T+1}/m - \delta_\rho)$-secure multiplex
network coding.
\end{theorem}

\begin{remark}
The condition (\ref{zerocond}) for almost zero mutual information
can become true for $\mu=1$ if
$\delta_\rho < n-\frac{k_{\mathcal{I}}}{m}-1$,
which is equivalent to $H_{1+\rho}(S_{\overline{\mathcal{I}}}|S_{\mathcal{I}}) / (m \ln q)> 1$.
A sufficient condition for (\ref{zerocond}) to hold for $\mu=1$
is that the conditional R\'enyi entropy of $S_{\overline{\mathcal{I}}}$
given $S_{\mathcal{I}}$ is $>\ln q$ for some $\rho$,
which is equivalent to $S_{\overline{\mathcal{I}}}$ has at least
one $\mathbf{F}_q$ symbol of conditional randomness given $S_{\mathcal{I}}$.
So we can see that the previous argument can ensure almost zero
mutual information with messages very far from independence and uniformity.
\end{remark}

\begin{remark}
The meaning of $C_1$ is as follows:
At Eqs.\ (\ref{eq1002}) and (\ref{eq1003}),
there might not exist a realization $\ell$ of $L$ that
satisfies Eqs.\ (\ref{eq1002}) and (\ref{eq1003})
for all subsets $\mathcal{I}$ of $\{1$, \ldots, $T\}$ simultaneously.
By sacrificing the tightness of the upper bounds, we
ensure the existence of $\ell$ satisfying Eqs.\ (\ref{eq:ub5})
and (\ref{eq:ub6}) for all $\mathcal{I}$.
\end{remark}

\begin{remark}\label{rem:zeromutual}
Under the assumption that all messages $S_1$, \ldots, $S_{T+1}$ are
uniform and independent,
the mutual information
can be made exactly zero for every eavesdropping matrix $B$.
The reason is as follows:
For fixed $B$ and $L=\ell$, we have
\begin{equation}
I(S_{\mathcal{I}};BLS^t|L=\ell)= H(S_{\mathcal{I}}|L=\ell) - H (S_{\mathcal{I}}|BLS^t,L=\ell).
\label{eq:mut}
\end{equation}
The first term $H(S_{\mathcal{I}}|L=\ell)$ is an integer multiple of
$\ln q$ since $S_{\mathcal{I}}$ is assumed to have the uniform
distribution. 
Let
$\alpha_{\mathcal{I}}$ be the projection from
$\prod_{i=1}^{T+1} \mathbf{F}_q^{k_i}$ to $\prod_{i\in\mathcal{I}} \mathbf{F}_q^{k_i}$
for $\emptyset \neq \mathcal{I} \subseteq \{1$, \ldots, $T\}$.
For fixed $B$ and $\ell$, and a given realization $z$ of $B\ell S^t$,
the set of solutions $s$ such that $z = B\ell s$ is written
as $\ker(B\ell) + $ some vector $v$.
This means that the set of
 possible candidates of $S_{\mathcal{I}}$ given
realization $z$ of $B\ell S^t$ is written as $\alpha_{\mathcal{I}}(\ker(B\ell)) +
\alpha_{\mathcal{I}}(v)$, and $S_{\mathcal{I}}$ given
realization $z$ is uniformly distributed on 
$\alpha_{\mathcal{I}}(\ker(B\ell)) +
\alpha_{\mathcal{I}}(v)$.
Since the cardinality of $\alpha_{\mathcal{I}}(\ker(B\ell)) +
\alpha_{\mathcal{I}}(v)$ is independent of $\ell S^t$ for fixed $B$ and $\ell$,
the second term $H (S_{\mathcal{I}}|BLS^t,L=\ell)$ is also an
integer multiple of $\ln q$. Therefore, if Eq.\ (\ref{eq:ub5})
holds for every $B$ as verified in Eq.\ (\ref{eq:prob})
and
the RHS of Eq.\ (\ref{eq:ub5}) is $<\ln q$, then
the LHS of Eq.\ (\ref{eq:ub5}) must be zero.
Observe that under this assumption our modified construction is
a universal $(0,k_{T+1}/m)$-secure multiplex network coding
in the exact sense of Definition \ref{def:univstrongsec}.
The parameter $k_{T+1}/m$ is optimal according to \cite{cai11survey}.
\end{remark}

\subsection{Evaluation of the required resource}
In this subsection, we evaluate the amount of required
resource in our proposal.
One can make convergence of Eq.\ (\ref{eq:ub5}) arbitrarily
slow by decreasing the difference between LHS and RHS of
Eq.\ (\ref{zerocond}), which makes evaluation of required size of $m$
very difficult.

To overcome the above difficulty,
we consider
$(\epsilon_I$, $k_{T+1}/m - \delta_\rho-\epsilon_\mu)$-secure
  multiplex network coding,
with which we have to ensure small mutual information only
for $\mu < n-k_{\mathcal{I}}/m - \delta_\rho - \epsilon_\mu$.
This assumption makes the difference between LHS and RHS of
Eq.\ (\ref{zerocond}) at least $\epsilon_\mu$, which enables us
to provide an upper bound on $m$.

\begin{proposition}\label{prop:evalm}
For given $n$, $q$, $T$, $\rho$, $\delta_\rho$, $\epsilon_I$,
$\epsilon_p$ and $\epsilon_\mu$,
\[
m \geq \frac{\frac{(n+1)^2}{4} + \log_q (2n (2^T-1)) - \log_q (\rho\epsilon_p\epsilon_I)}{\rho\epsilon_\mu}
\]
is sufficient to ensure
that a random choice of $L$ gives an
$(\epsilon_I$, $k_{T+1}/m - \delta_\rho-\epsilon_\mu)$-secure multiplex
network coding with probability at least $1-\epsilon_p$.
\end{proposition}

\begin{IEEEproof}
By Eq.\ (\ref{eq:prob}) we have to
choose $C_1$ with
\begin{equation}
C_1 \geq 2 \times (2^T-1)\times  n q^{\frac{(n+1)^2}{4}}/\epsilon_p. \label{eq10001}
\end{equation}
By Eq.\ (\ref{eq1004}), to make the mutual information $\leq \epsilon_I$,
we see
\begin{equation}
\ln \frac{C_1}{\rho} - m \rho \epsilon_\mu \ln q \leq \ln \epsilon_I\label{eq10002}
\end{equation}
is sufficient.
The condition (\ref{eq10002}) is equivalent to
\begin{eqnarray*}
&&m \geq (\ln\frac{2 \times (2^T-1)\times  n q^{\frac{(n+1)^2}{4}}}{\rho\epsilon_p\epsilon_I})/(\rho\epsilon_\mu\ln q)\\
&\Leftrightarrow&
m \geq \frac{\frac{(n+1)^2}{4} + \log_q (2n (2^T-1)) - \log_q (\rho\epsilon_p\epsilon_I)}{\rho\epsilon_\mu}
\end{eqnarray*}
\end{IEEEproof}

We comment on the required field size and the computational complexity
of code construction of our proposal
and previous proposals realizing the security.
The proposed construction works with any given field size $q$,
as well as \cite{kurihara15,silva09}.
The required sizes of $q$ in \cite{bhattad05,harada08}
are not explicitly given but they seem quite large.

Instead of increasing $q$,
we need to increase $m$ to satisfy the maximum allowable
mutual information to the eavesdropper,
as shown in Proposition \ref{prop:evalm}.
Proposition \ref{prop:evalm} indicates that
a small value of $\epsilon_\mu$ makes the required size of $m$
large, because smaller $\epsilon_\mu$ makes
the convergence of Eq.\ (\ref{eq:ub5}) slower.
In \cite{silva09}, $m \geq n$ is sufficient for explicit construction
of a code, and in \cite{kurihara15} $m \geq 2n$ is sufficient,
while neither \cite{kurihara15,silva09} realizes almost zero mutual information
with dependent or non-uniform multiple messages.

The complexity of code construction of our proposal is
$m^2n^2$ because of the random choice of $mn\times mn$ matrix.
The codes  in \cite{kurihara15,silva09}
are the Gabidulin codes \cite{Gabidulin1985}
of length $n$ over $\mathbf{F}_{q^m}$ and
construction of an encoding matrix at the
source node can be done in $m^2n^2$ arithmetic
operations in $\mathbf{F}_q$.
We note that for small $\epsilon_\mu$ the required size of $m$
in our proposal can be much larger than \cite{kurihara15,silva09}.
The complexities of code constructions in \cite{bhattad05,harada08}
are not given but they seem quite large.

\subsection{Numerical example of explicit computation of required block size $m$}\label{sec34}
In this section we give a numerical example of computing
required block length $m$ in order to ensure the mutual
information is below some value.
In order to do so,
we need an estimate of $\mathbf{E}[P_{S_{\overline{\mathcal{I}}}|S_{\mathcal{I}}}(S_{\overline{\mathcal{I}}}|S_{\mathcal{I}})^\rho]$.
We assume to have $\delta_{0.5}=0.5$ in Eq.\ (\ref{eq:delta0})
at $\rho=0.5$.

Let $q=256$, $n=10$, $\mu=3$, $T=5$, $k_i = 2m$ for all $i$.
We do not have $S_{T+1}$.
We want to ensure that we choose $\ell$ with
probability at least $1-10^{-12}$ such that $I(S_i; BLS^t|L=\ell) < 10^{-6}$
for all $i=1$, \ldots, $5$.
By Eq.\ (\ref{eq:prob}) we choose $C_1$ as
\begin{eqnarray*}
&&2 \times n q^{\frac{(n+1)^2}{4}} (2^T-1)/C_1 = 10^{-12}\\
&\Leftrightarrow &
C_1 = 2 \times 10 \times 256^{11^2/4}(2^T-1) 10^{12} 
\end{eqnarray*}

By using $\delta_\rho$, we can upper bound the RHS of Eq.\ (\ref{eq:ub5})
as follows:
\begin{eqnarray}
&&C_1 q^{m \rho \mu}\mathbf{E}[P_{S_{\overline{\mathcal{I}}}|S_{\mathcal{I}}}(S_{\overline{\mathcal{I}}}|S_{\mathcal{I}})^\rho]/\rho\nonumber\\
&=& C_1 \exp_q (m \rho (\mu + \frac{H_{1+\rho}(S_{\overline{\mathcal{I}}}|S_{\mathcal{I}})}{m\ln q})/\rho\nonumber\\
&\leq& C_1 \exp_q (m \rho(\mu - n + k_{\mathcal{I}}/m + \delta_\rho))/\rho
\mbox{ (by Eq.\ (\ref{eq:delta0}))}. \label{eq1006}
\end{eqnarray}
In order to keep the above upper bound to be below $10^{-6}$ we
have to choose
\begin{eqnarray*}
&& C_1 \exp_q (m \rho(\mu - n + k_{\mathcal{I}}/m + \delta_\rho))/\rho < 10^{-6}\\
&\Leftrightarrow & m > -\frac{\log_q (10^6 C_1/\rho)}{\rho (\mu - n + k_{\mathcal{I}}/m + \delta_{\rho})}\\
&\Leftrightarrow & m > -\frac{\log_{256} (10^6 \times 2 \times 10 \times 256^{121/4}(2^5-1)10^{12}/0.5)}{0.5(3 - 10 + 2 +0.5)}\\
&\Leftarrow & m \geq 
17.3373
\end{eqnarray*}
This means that we can choose $m=18$ and
should choose the matrix $L$ at least as large as
$180 \times 180$
over $\mathbf{F}_{256}$, which is implementable.
Recall that we assumed $n=10$ outgoing (logical) links from
the source node and that each outgoing link carries
$m=18$ symbols in single coding block in this example.
We note that the above computation corresponds to the case
$\epsilon_I = 10^{-6}$, $\epsilon_p = 10^{-12}$ and $\epsilon_\mu = 4.5$
in Proposition \ref{prop:evalm},
and realizes $(10^{-6}$, $-5)$-secure
  multiplex network coding in the sense of
Definition \ref{def:univstrongsec} with probability $1-10^{-12}$.
Relatively small $m$ comes from the choice of $\epsilon_\mu = 4.5$.
If we want to realize the same level of security for any
triple of $S_1$, \ldots $S_5$ instead of single $S_i$,
then $\epsilon_\mu$ becomes  $0.5$ and the required size of $m$ becomes $9 (=4.5/0.5)$
times larger than this example, which realizes
$(10^{-6}$, $-1)$-secure multiplex network coding.
Since $\delta_\rho >0$, we cannot realize
$(\epsilon_p$, $0)$-secure multiplex network coding
without use of the dummy message $S_{T+1}$, which is not
used in this example.Use of the dummy message
$S_{T+1}$ also decreases the required size of $m$.

\begin{remark}
A vector in $\mathbf{F}_q^{mn}$ can be identified with an element
in $\mathbf{F}_{q^{mn}}$, and multiplication by a nonzero element
in $\mathbf{F}_{q^{mn}}$ is an $\mathbf{F}_q$-linear mapping and
can be identified with an element in $\mathcal{L}$.
Let $\mathcal{L}_{\mathbf{F}_{q^{mn}}}$ be a commutative
subgroup of $\mathcal{L}$ whose elements can be identified with
nonzero elements in $\mathbf{F}_{q^{mn}}$.
By looking at the proof of Lemma \ref{lem10} in Appendix \ref{app:b},
we can see that $\mathcal{L}_{\mathbf{F}_{q^{mn}}}$ can be used in place of
$\mathcal{L}$ in our modified construction.
Necessary storage space to record choice of an element in 
$\mathcal{L}_{\mathbf{F}_{q^{mn}}}$ is that of $mn$ $\mathbf{F}_q$ symbols
and is smaller than that of 
$\mathcal{L}$.
Matrix multiplication by an element in 
$\mathcal{L}_{\mathbf{F}_{q^{mn}}}$ is at least as fast as that in 
$\mathcal{L}$.
\end{remark}

\section{Conclusion}\label{sec4}
In the secure network coding,
there was loss of information rate due to inclusion of random bits at the source node.
Weakly and strongly secure network coding \cite{bhattad05,cai09secure,harada08,silva09} remove that loss of information rate
by using multiple messages to be kept secret from an eavesdropper,
which require huge computational complexity in code construction or
huge finite field size.
In addition to this, the previous studies assumed uniform and independent
multiple messages, which seems too strong assumption in practice.
In this paper, we have shown that random linear transform of
multiple messages at the source node realizes the strongly
secure (called secure multiplex network coding in this paper)
network coding with arbitrary high probability
with sufficiently large block length.
We did not assume uniformity nor independence in multiple messages.
Our numerical example in Section \ref{sec34}
showed that ``sufficiently large block length'' can
be small.
We studied the secure network coding
from separately the source coding of messages.
Joint source and network coding
might improve the performance, but we leave the study of
such a joint encoding as a future.

\appendices
\section{Proof of Proposition \ref{thm2}}\label{app:a}
In order to show Proposition \ref{thm2},
we introduce the following lemma.
\begin{lemma}\label{lem1}
Under the same assumption as Proposition \ref{thm2},
we have
\begin{equation}
\mathbf{E}_f [\exp(-\rho H(F(A_1)|A_2,F=f))]  \leq   
|\mathcal{A}_3|^{-\rho} + \mathbf{E}[P_{A_1|A_2}(A_1|A_2)^\rho]\label{eq:lem1}
\end{equation}
for $0\leq \rho \leq 1$.
\end{lemma}

\begin{IEEEproof}[Proof of Proposition \ref{thm2}]
\begin{align*}
&  \mathbf{E}_f [\exp(\rho I(F(A_1);A_2|F=f))]\\
&= \mathbf{E}_f [\exp(\rho \underbrace{H(F(A_1)|F=f)}_{\leq \log_q |\mathcal{A}_3|} - \rho H(F(A_1)|A_2,F=f))]\\
&\leq \mathbf{E}_f [|\mathcal{A}_3|^\rho \exp(-\rho H(F(A_1)|A_2,F=f))]\\
&\leq |\mathcal{A}_3|^\rho (|\mathcal{A}_3|^{-\rho} + \mathbf{E}[P_{A_1|A_2}(A_1|A_2)^\rho]) \textrm{ (by Eq. (\ref{eq:lem1}))}\\
&= 1+ |\mathcal{A}_3|^\rho\mathbf{E}[P_{A_1|A_2}(A_1|A_2)^\rho].
\end{align*}
\end{IEEEproof}

\begin{IEEEproof}[Proof of Lemma \ref{lem1}]
Fix $a_2 \in \mathcal{A}_2$.
The concavity of $x^\rho$ for $0\leq\rho\leq 1$ implies 
\begin{align}
& \mathbf{E}_f \Bigl[\sum_{a_3 \in \mathcal{A}_3} P_{f(A_1)|A_2}(a_3|a_2)^{1+\rho}\Bigr] \nonumber\\
&= \mathbf{E}_f \Bigl[\sum_{a_3 \in \mathcal{A}_3}
\underbrace{P_{f(A_1)|A_2}(a_3|a_2)}_{=\sum_{a_1 \in f^{-1}(a_3)} P_{A_1|A_2}(a_1|a_2)}
P_{f(A_1)|A_2}(a_3|a_2)^{\rho}\Bigr] \nonumber\\
&= \mathbf{E}_f \Bigl[\sum_{a_1\in \mathcal{A}_1} P_{A_1|A_2}(a_1|a_2)  \sum_{a_1'\in f^{-1}(f(a_1))} P_{A_1|A_2}(a_1'|a_2)^{\rho}\Bigr]\nonumber\\
&= \sum_{a_1\in \mathcal{A}_1} P_{A_1|A_2}(a_1|a_2) \mathbf{E}_f \Bigl[ \sum_{a_1'\in f^{-1}(f(a_1))} P_{A_1|A_2}(a_1'|a_2)^\rho\Bigr]\nonumber\\
&\leq  \sum_{a_1\in \mathcal{A}_1} P_{A_1|A_2}(a_1|a_2) \Bigl(\underbrace{\mathbf{E}_f  \Bigl[\sum_{a_1'\in f^{-1}(f(a_1))} P_{A_1|A_2}(a_1'|a_2)\Bigr]}_{(**)}\Bigr)^\rho. \label{eq200}
\end{align}
For a fixed realization $a_2$ of $A_2$,
by the assumption in Proposition \ref{thm2} two random variables
$F$ and $A_1$ are statistically independent,
which implies the distribution of $f$ in (**) is independent of
$a_1$.
Since $f$ is chosen from a family of two-universal hash functions
defined in Definition \ref{def:twouniv}, we have $P(
a_1'\in F^{-1}(F(a_1))\setminus\{a_1\}) \leq 1/|\mathcal{A}_3|$ for
$a_1 \neq a'_1 \in \mathcal{A}_1$ and
\begin{align*}
(**) & =\mathbf{E}_f  \Bigl[P_{A_1|A_2}(a_1|a_2) + \sum_{a_1'\in f^{-1}(f(a_1))\setminus\{a_1\}} P_{A_1|A_2}(a_1'|a_2)\Bigr]\\
&\leq P_{A_1|A_2}(a_1|a_2) + \sum_{a_1 \neq a_1' \in \mathcal{A}_1} \frac{P_{A_1|A_2}(a_1'|a_2)}{|\mathcal{A}_3|}\\
&\leq P_{A_1|A_2}(a_1|a_2) + |\mathcal{A}_3|^{-1}.
\end{align*}
Since any two positive numbers $x$ and $y$ satisfy
$(x+y)^\rho \leq x^\rho + y^\rho$ for $0 \leq \rho \leq 1$,
we have
\begin{equation}
(P_{A_1|A_2}(a_1|a_2) + |\mathcal{A}_3|^{-1})^\rho \leq P_{A_1|A_2}(a_1|a_2)^\rho + |\mathcal{A}_3|^{-\rho}.
\label{eq201}
\end{equation}
By Eqs.\ (\ref{eq200}) and (\ref{eq201}) we can see
\[
\mathbf{E}_f \Bigl[\sum_{a_3 \in \mathcal{A}_3} P_{f(A_1)|A_2}(a_3|a_2)^{1+\rho}\Bigr] \leq
\sum_{a_1 \in \mathcal{A}_1} P_{A_1|A_2}(a_1|a_2)^{1+\rho} + |\mathcal{A}_3|^{-\rho}.
\]
Taking the average over $A_2$ of the both sides
of the last equation, we have
\begin{equation}
\mathbf{E}_f [\mathbf{E}_{A_1A_2} [P_{f(A_1)|A_2}(f(A_1)|A_2)^{\rho}]] \leq
\mathbf{E}_{A_1A_2}  [P_{A_1|A_2}(A_1|A_2)^{\rho}] + |\mathcal{A}_3|^{-\rho}. \label{eq203}
\end{equation}
Define $g(\rho) = \mathbf{E}_{A_1A_2} [P_{f(A_1)|A_2}(f(A_1)|A_2)^{\rho}]$
as a function of $\rho$ with fixed $f$ and $P_{A_1A_2}$, and
$h(\rho) = \ln g(\rho)$.
We have
\begin{align*}
g'(\rho)
&= \mathbf{E}_{A_1A_2} [P_{f(A_1)|A_2}(f(A_1)|A_2)^{\rho} \ln P_{f(A_1)|A_2}(f(A_1)|A_2)],\\
g''(\rho)
&= \mathbf{E}_{A_1A_2} [P_{f(A_1)|A_2}(f(A_1)|A_2)^{\rho} (\ln P_{f(A_1)|A_2}(f(A_1)|A_2))^2],\\
h'(\rho) &= g'(\rho)/g(\rho),\\
h''(\rho) &= \frac{g''(\rho)g(\rho) - [g'(\rho)]^2}{g(\rho)^2}.
\end{align*}
Define $(A_1'$, $A_2')$ to be the random variables that have the same
joint distribution as $(A_1,A_2)$ and statistically independent of
$A_1$ and $A_2$.
To examine the sign of $h''(\rho)$ we compute 
\begin{align*}
&  g''(\rho)g(\rho) - [g'(\rho)]^2\\
&= \mathbf{E}_{A_1A_2A_1'A_2'} [P_{f(A_1)A_2}(f(A_1),A_2)^{\rho}P_{f(A_1)A_2}(f(A_1'),A_2')^{\rho}  \\*
&\qquad \{(\ln P_{f(A_1)|A_2}(f(A_1)|A_2))^2  \\
&\qquad - \ln P_{f(A_1)|A_2}(A_1|A_2)\ln P_{f(A_1)|A_2}(A_1'|A_2')\}]\\
&= \frac{1}{2} \mathbf{E}_{A_1A_2A_1'A_2'} [P_{f(A_1)A_2}(f(A_1),A_2)^{\rho}P_{f(A_1)A_2}(f(A_1'),A_2')^{\rho}
\\*
&\qquad\{(\ln P_{f(A_1)|A_2}(f(A_1)|A_2))^2
+ (\ln P_{f(A_1)|A_2}(f(A_1')|A_2'))^2 \\*
&\qquad - 2\ln P_{f(A_1)|A_2}(f(A_1)|A_2)\ln P_{f(A_1)|A_2}(f(A_1')|A_2')\}]\\
&= \frac{1}{2} \mathbf{E}_{A_1A_2A_1'A_2'} [P_{f(A_1)A_2}(f(A_1),A_2)^{\rho}P_{f(A_1)A_2}(f(A_1'),A_2')^{\rho}\\*
&\qquad \{\ln P_{f(A_1)|A_2}(f(A_1)|A_2) - \ln P_{f(A_1)|A_2}(f(A_1')|A_2')\}^2]\\
&\geq 0.
\end{align*}
This means that $h''(\rho) \geq 0$ and $h(\rho)$ is convex.
We can see 
\begin{align}
\mathbf{E}_{A_1A_2} [P_{f(A_1)|A_2}(f(A_1)|A_2)^{\rho}] &= \exp(h(\rho)) \nonumber\\
&\geq  \exp(\underbrace{h(0)}_{=0} + \rho h'(0))\nonumber\\
&= \exp(-\rho H(f(A_1)|A_2)). \label{eq204}
\end{align}
By Eqs.\ (\ref{eq203}) and (\ref{eq204}) we see that
Eq.\ (\ref{eq:lem1}) holds.
\end{IEEEproof}

\section{Proof of Lemma~\ref{lem10}}\label{app:b}
We shall prove Lemma \ref{lem10} in this Appendix.
Let $\mathcal{L}$ be a subgroup of the group of all bijective
linear maps on $\mathbf{F}_q^{mn}$.
For $\vec{x} \in \mathbf{F}_q^{mn}$,
the orbit $O(\vec{x})$ of $\vec{x}$ under the action of $\mathcal{L}$
is defined by
\[
O(\vec{x}) = \{ L\vec{x} \mid L \in \mathcal{L}\}.
\]
\begin{lemma}\label{lem:orbit0}
Let $\vec{x}$, $\vec{y}$ be two different vectors
belonging to $O(\vec{z})$.
We have
\[
|\{ L \in \mathcal{L} \mid L\vec{z}=\vec{x} \}|
=
|\{ L \in \mathcal{L} \mid L\vec{z}=\vec{y} \}|.
\]
\end{lemma}
\begin{IEEEproof}
Let $K\in \mathcal{L}$ such that $K\vec{x}=\vec{y}$. We have
\begin{eqnarray*}
&& |\{ L \in \mathcal{L} \mid L\vec{z}=\vec{x} \}|\\
&=& |\{ L \in \mathcal{L} \mid KL\vec{z}=K\vec{x} \}|\\
&=& |\{ L \in \mathcal{L} \mid KL\vec{z}=\vec{y} \}|\\
&=& |\{ L \in \mathcal{L} \mid L\vec{z}=\vec{y} \}|.
\end{eqnarray*}
\end{IEEEproof}

\begin{lemma}\label{lem:orbit}
Let $B$ be an $m \mu \times mn$ matrix,
$\ker(B) = \{ \vec{x} \in \mathbf{F}_q^{mn} \mid
B\vec{x} = \vec{0} \}$, and
$\mathrm{im}(B) = \{ B\vec{x} \mid \vec{x} \in \mathbf{F}_q^{mn}\}$.
The family of functions $\{ B L \mid L \in \mathcal{L}\}$
with uniformly distributed $L$
is a family of two-universal hash functions from $\mathbf{F}_q^{mn}$
to $\mathrm{im}(B)$ if and only if
\[
\frac{|O(\vec{v}) \cap \ker(B)|}{|O(\vec{v})|}
\leq \frac{1}{|\mathrm{im}(B)|}
\]
for all $\vec{v} \in \mathbf{F}_q^{mn}\setminus \{\vec{0}\}$.
\end{lemma}

\begin{IEEEproof}
With the uniform distribution on $\mathcal{L}$,
LHS of Eq.\ (\ref{eq:two}) is equal to
\begin{eqnarray*}
&&\frac{|\{ L \in \mathcal{L} \mid BL\vec{x}_1=BL\vec{x}_2\}|}{|\mathcal{L}|}\\
&=&\frac{|\{ L \in \mathcal{L} \mid BL(\vec{x}_1-\vec{x}_2) =\vec{0}\}|}{|\mathcal{L}|}\\
&=&\frac{|\{ L \in \mathcal{L} \mid L(\vec{x}_1-\vec{x}_2) \in \ker(B) \}|}{|\mathcal{L}|}\\
&=&\frac{|\{ L \in \mathcal{L} \mid L(\vec{x}_1-\vec{x}_2) \in O(\vec{x}_1-\vec{x}_2)\cap\ker(B)\}|}{|\{ L \in \mathcal{L} \mid L(\vec{x}_1-\vec{x}_2) \in O(\vec{x}_1-\vec{x}_2)\}|}\\
&=&\frac{|O(\vec{x}_1-\vec{x}_2)\cap \ker(B)|}{|O(\vec{x}_1-\vec{x}_2)|}
\mbox{ (by Lemma~\ref{lem:orbit0})}.
\end{eqnarray*}
Renaming $\vec{x}_1-\vec{x}_2$ to $\vec{v}$ proves the lemma.
\end{IEEEproof}

\begin{proposition}\label{prop:app}
If $\mathcal{L}$ is the set of all bijective linear maps on
$\mathbf{F}_q^{mn}$, then
$\{ B L \mid L \in \mathcal{L}\}$ with uniformly distributed $L$
is a family of two-universal hash functions
from $\mathbf{F}_q^{mn}$ to $\mathrm{im}(B)$.
\end{proposition}
\begin{IEEEproof}
For a nonzero $\vec{v} \in \mathbf{F}_q^{mn}$,
we have $O(\vec{v}) = \mathbf{F}_q^{mn}\setminus\{\vec{0}\}$,
which implies
\begin{eqnarray*}
&&|O(\vec{v})| = |\mathbf{F}_q^{mn}|-1,\\
&&|O(\vec{v}) \cap \ker(B)|=\frac{|\mathbf{F}_q^{mn}|}{|\mathrm{im}(B)|} - 1.
\end{eqnarray*}
By Lemma \ref{lem:orbit} we can see that the proposition is true.
\end{IEEEproof}

\begin{IEEEproof}[Proof of Lemma \ref{lem10}]
Lemma \ref{lem10} is equivalent to Proposition \ref{prop:app}.
\end{IEEEproof}

\section*{Acknowledgment}
The authors thank anonymous reviewers of NetCod 2011 and
this journal
for
carefully reading the previous manuscripts and pointing out
their shortcomings.
The first author would like to thank Prof.\ H.\  Yamamoto to teach him
the secure multiplex coding,
Prof.\ S.\ Watanabe to point out the relation between
the proposed scheme and \cite{harada08},
Dr.\ J.\ Kurihara to point out the relation between
the proposed scheme and \cite{silva09},
Dr.\ J.\ Muramatsu and Prof.\ T.\ Ogawa for the helpful discussion on the universal
coding.
A part of this research was done during the first author's stay
at the Institute of Network Coding, the Chinese University
of Hong Kong, and Department of Mathematical Sciences, Aalborg University.
He greatly appreciates the hospitality by Prof.\ R.\ Yeung and
Prof.\ O.\ Geil.


\begin{IEEEbiographynophoto}{Ryutaroh Matsumoto}
(M'00) was born in Nagoya, Japan, on November 29, 1973. He
received the B.E. degree in computer science, the M.E. degree in
information processing, and the Ph.D. degree in electrical and
electronic engineering, all from Tokyo Institute of Technology, Japan,
in 1996, 1998 and 2001, respectively. He was an Assistant Professor
from 2001 to 2004, and an Associate Professor from  2004 2017 in
the Department of Information and Communications Engineering, Tokyo
Institute of Technology.
He has been an Associate Professor
in the Department of Information and Communication Engineering,
Nagoya University since April 2017.
He also served as a Velux Visiting Professor
at the Department of Mathematical Sciences, Aalborg University,
Denmark, in 2011 and 2014.  His research interests include
error-correcting codes, quantum information theory, information
theoretic security, and communication theory. Dr. Matsumoto received
the Young Engineer Award from IEICE and the Ericsson Young Scientist
Award from Ericsson Japan in 2001. He received the Best Paper Awards
from IEICE in 2001, 2008, 2011 and 2014.
\end{IEEEbiographynophoto}

\begin{IEEEbiographynophoto}{Masahito Hayashi}(M'06--SM'13--F'17) was born in Japan in 1971.
He received the B.S. degree from the Faculty of Sciences in Kyoto
University, Japan, in 1994 and the M.S. and Ph.D. degrees in Mathematics from
Kyoto University, Japan, in 1996 and 1999, respectively. He worked in Kyoto University as a Research Fellow of the Japan Society of the
Promotion of Science (JSPS) from 1998 to 2000,
and worked in the Laboratory for Mathematical Neuroscience,
Brain Science Institute, RIKEN from 2000 to 2003,
and worked in ERATO Quantum Computation and Information Project,
Japan Science and Technology Agency (JST) as the Research Head from 2000 to 2006.
He also worked in the Superrobust Computation Project Information Science and Technology Strategic Core (21st Century COE by MEXT) Graduate School of Information Science and Technology, The University of Tokyo as Adjunct Associate Professor from 2004 to 2007.
He worked in the Graduate School of Information Sciences, Tohoku University as Associate Professor from 2007 to 2012.
In 2012, he joined the Graduate School of Mathematics, Nagoya University as Professor.
He also worked in Centre for Quantum Technologies, National University of Singapore as Visiting Research Associate Professor from 2009 to 2012
and as Visiting Research Professor from 2012 to now.
In 2011, he received Information Theory Society Paper Award (2011) for ``Information-Spectrum Approach to Second-Order Coding Rate in Channel Coding''.
In 2016, he received the Japan Academy Medal from the Japan Academy
and the JSPS Prize from Japan Society for the Promotion of Science.

In 2006, he published the book ``Quantum Information: An Introduction''  from Springer, whose revised version was published as ``Quantum Information Theory: Mathematical Foundation'' from Graduate Texts in Physics, Springer in 2016.
In 2016, he published other two books ``Group Representation for Quantum Theory'' and ``A Group Theoretic Approach to Quantum Information'' from Springer.
He is on the Editorial Board of {\it International Journal of Quantum Information}
and {\it International Journal On Advances in Security}.
His research interests include classical and quantum information theory and classical and quantum statistical inference.
\end{IEEEbiographynophoto}

\end{document}